\documentclass{article}
\usepackage{fullpage,graphicx}
\usepackage{rotating}
\usepackage{array}
\usepackage{color}
\usepackage{url}
\usepackage{cite}
\usepackage{subfigure}
\usepackage[]{algorithmic} 
\usepackage[]{algorithm,caption}
\DeclareGraphicsExtensions{.pdf,.jpeg,.png}

\author{
	Huda Al-Nayyef$^{1, 2}$, Christophe Guyeux$^1$, and	Jacques M. Bahi$^1$\\
	$^1$~FEMTO-ST Institute, UMR 6174 CNRS, DISC Computer Science Department \\
	Universit\'e de Franche-Comt\'e,
	16, Rue de Gray, 25000 Besan\c{c}on, France\\
$^2$~Computer Science Department, University of Mustansiriyah, Iraq\\
\{huda.al-nayyef, christophe.guyeux, jacques.bahi\}@univ-fcomte.fr
}

\title{A Pipeline for Insertion Sequence Detection and Study for Bacterial Genome}

\begin{document}
\maketitle

\begin{abstract}
Insertion Sequences (ISs) are small DNA segments that have the ability of moving themselves into genomes.
These types of mobile genetic elements (MGEs) seem to play an essential role in genomes rearrangements and evolution of prokaryotic genomes, but the tools that deal with discovering ISs in an efficient and accurate way are still too few and not totally precise. Two main factors have big effects on IS discovery, namely: genes annotation and functionality prediction. Indeed, some specific genes called ``transposases'' are enzymes that are responsible of the production and catalysis for such transposition, but there is currently no fully accurate method that could decide whether a given predicted gene is either a real transposase or not. This is why authors of this article aim at designing a novel pipeline for ISs detection and classification, which embeds the most recently available tools developed in this field of research, namely OASIS (Optimized Annotation System for Insertion Sequence) and ISFinder database (an up-to-date and accurate repository of known insertion sequences). As this latter depend on predicted coding sequences, the proposed pipeline will encompass too various kinds of bacterial genes annotation tools (that is, Prokka, BASys, and Prodigal). A complete IS detection and classification pipeline is then proposed and tested on a set of 23 complete genomes of \emph{Pseudomonas aeruginosa}.
This pipeline can also be used as an investigator of annotation tools performance, which has led us to conclude that Prodigal is the best software for IS prediction. A deepen study regarding IS elements in \emph{P.aeruginosa} has then been conducted, leading to the conclusion that close genomes inside this species have also a close numbers of IS families and groups.
\end{abstract}

\section{Introduction}

The number of completely sequenced bacterial and archaeal genomes are rising steadily, 
such an increasing makes it possible to develop novel kind of large scale approaches 
to understand genomes structure and evolution over time. Gene content prediction and genome comparison have both provided new important information and 
deciphering keys to understand evolution of prokaryotes ~\cite{varani2011issaga}.
 Important sequences in understanding rearrangement of genomes during evolution are so-called 
transposable elements (TEs), which are DNA fragments or segments that 
have the ability to insert themselves into new chromosomal locations, and often 
make duplicate copies of themselves during transposition process~\cite{nref1hawkins2006differential}. Remark that, in 
bacterial reign, only cut-and-paste mechanism of transposition can be found, the 
transposable elements involved in such a move being the insertion sequences (ISs).

\begin{figure}[ht!]
\centering
\includegraphics[scale=1]{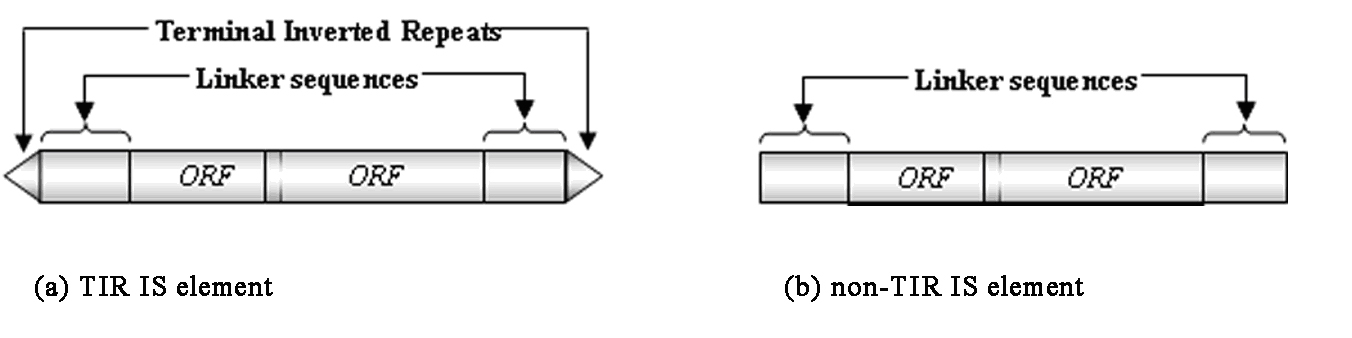}
\caption{IS element types~\cite{zhou2008insertion}}
\label{ISfig}
\end{figure}

Insertion sequences 
range in size from 600 to more than 3000 
bp.  They are divided into 26 main different families in prokaryotes, as described in 
ISFinder\footnote{\url{www-is.biotoul.fr}}~\cite{siguier2006isfinder}, an international 
reference database for bacterial and archaeal ISs that includes background information on  
transposons.  The main function of ISFinder is to assign IS names and to produce a focal 
point for a coherent nomenclature for all discovered insertion sequences. 
This database includes over than 3500 bacterial ISs~\cite{emma6hickman2010integrating,zhou2008insertion}. 
Data come from a detection of repeated patterns, which can be 
easily found by using homology-based techniques~\cite{chr4feschotte2007computational}. 
Classification process of families, for its part, depends on transposases homology 
and overall genetic organization. Indeed, most ISs consist of short inverted repeat 
sequences that flank one or more open reading frames (ORFs, see Figure~\ref{ISfig}), 
whose products encode the transposase proteins necessary for transposition process. 
The main problem with such approaches for ISs detection and classification
is that they are obviously highly dependent on
the annotations, and existing tools evoked above only use 
the NCBI ones, whose quality is limited and very variable.

In this research work, the authors' intention is to find an accurate method for discovering insertion sequences in prokaryotic genomes. 
To achieve this goal, we propose to use one of the most recent computational tool 
for automated annotation of insertion sequences, namely OASIS, together with the 
international database for all known IS sequences (ISFinder). 
More precisely, OASIS works with genbank files that have fully 
described genes functionality: this tool identifies ISs in 
each genome by finding conserved regions surrounding already-annotated 
transposases. Such technique makes it possible to discover new insertion sequences, even 
if they are not in ISFinder database.
A novel pipeline that solves the dependence on NCBI annotations, 
and that works with any annotation tool (with or without description of 
gene functionality) is then proposed. 
The output of our pipeline contains all detected IS sequences supported with other important information like inverted repeats (IRs) sequences, lengths, positions, names of family and group, and other details that help in studying IS structures.

The contributions of this article can be summarized as follows. (1) 
A pipeline for insertion sequences discovery and classification is proposed, which does not depend 
on NCBI annotations. It uses unannotated genomes and embeds various annotation tools 
specific to Bacteria (such as Prokka, BASys, and Prodigal) in its
process. (2) Overlapping and consensus problems that naturally appear after 
merging annotation methods recalled above are solved, in order to obtain large and accurate number of ISs with their names of families and groups. And finally (3) the pipeline is tested 
on a set of 23 complete genomes of \emph{Pseudomonas aeruginosa}, and
biological consequences are outlined.

\color{black}
The remainder of this article is organized as follows. In Section~\ref{sec:state}, various tools 
for discovering IS elements in different species of Bacteria and Archaea are presented. 
The suggested methodology for increasing both the number and accuracy of detecting 
IS elements is explained in Section~\ref{sec:methods}. The pipeline is detailed in 
Section~\ref{sec:pip}, while an application example using 23 completed genomes of 
\emph{P. aeurigonsa} is provided in Section~\ref{sec:result}. 
This article ends by a conclusion section, in which the contributions are summarized and 
intended future work is detailed.

\section{State of the art in ISs detection or annotation}
\label{sec:state}

The study on the plant-pathogenic prokaryote \emph{Xanthomonas oryzae pv. oryzae (Xoo)}, which causes 
bacterial blight (one of the most important diseases of rice) was published in 2005
by Ochiai \emph{et al.}~\cite{art1ochiai2005genome}. They used GeneHacker~\cite{yada1996detection}, GenomeGambler version 1.51, and Glimmer program~\cite{delcher1999improved} for coding sequence prediction. 
Insertion sequences were finally classified by a BLAST analysis using ISFinder database evoked previously. 

\textbf{IScan}, developed by Wagner \emph{et al.}~\cite{emma2Wagner01082007}, has then 
been proposed in 2007. Inverted repeats are found using smith waterman local 
alignments on transposase references found with BLAST and used 
as a local database. This tool has been 
applied on 438 completely sequenced bacterial genomes by using BLAST with referenced
transposases, to determine which transposases are related to insertion sequences.
Touchon \emph{et al.}, for their parts, have analyzed  262 different 
bacterial and archaeal genomes downloaded from GenBank NCBI 
in 2007~\cite{art4touchon2007causes}. A coding sequence has then been considered as an IS element if its 
BLASTP best hit in ISFinder database  has an e-value lower than $10^{-10}$.

\textbf{ISA} has been created by Zhou \emph{et al.} in 2008~\cite{zhou2008insertion}. 
This annotation program depends on both NCBI annotations and ISFinder.
More precisely, authors manually collected 1,356 IS elements with both sequences
and terminal signals from the ISFinder database, which have
been used as templates for identification of all IS elements and map 
construction in the targeted genomes.
ISA, which is not publicly available, has finally been used for an analysis of
19 cyanobacterial and 31 archaeal annotated genomes downloaded from NCBI.  

In 2010, Plague \emph{et al.} analyzed the neighboring gene
orientations (NGOs) of all ISs in 326 fully sequenced bacterial chromosomes. They obtained primary annotations from the Comprehensive Microbial Resource database (release 1.0-20.0) at the Institute for Genomic 
Research\footnote{\url{http://cmr.tigr.org/tigr-scripts/CMR/CmrHomePage.cgi}}. 
Their approach for extracting IS elements from these genomes was to consider 
that a coding sequence with a best BLASTX hit e-value lower than $10^{-10}$ is 
an insertion sequence~\cite{art2plague2010intergenic}.
\textbf{ISsage}, for its part, has been developed in 2011 by Varani 
\emph{et al.}~\cite{varani2011issaga}. They used eight different bacterial genomes downloaded from NCBI, and produced a web application pipeline that allows semi-automated annotation 
based on BLAST against the ISFinder database. However ISsage cannot 
automatically identify new insertion sequences which are not already present in 
ISFinder database.

A new computational tool for automated annotation of ISs has then been 
released in 2012 by Robinson \emph{et al.}~\cite{robinson2012oasis}.
This tool has been called \textbf{OASIS}, which stands for ``Optimized 
Annotation System for Insertion Sequences''. They worked with 1,737 bacterial and archaeal genomes downloaded from NCBI. 
OASIS identifies ISs in each genome by finding conserved regions surrounding already-annotated transposase genes. OASIS uses a maximum likelihood algorithm 
to determine the edges of multicopy ISs based on conservation between their
surrounding regions. For defining inverted repeats, the same strategy as IScan 
was used (Smith-Waterman alignment). Authors also used hierarchical 
agglomerative
clustering to identify groups of IS lengths. The ISs set is then classified 
according to the family and group after a BLASTP best hit in ISFinder database 
with an e-value lower than $10^{-12}$. When a cluster 
cannot match with any entry of the database, the IS set is 
considered as new. Thus OASIS has the ability to discover new 
insertion sequences, that is, which cannot be found in ISFinder.
	
\begin{table}[H]
\center
\caption{Input set of 23 complete genomes of \emph{P. aeruginosa}}
\scalebox{0.7}{%
\begin{tabular}{|c|c|c|c|c|c|}
\hline
\textbf{} & \textbf{} & \textbf{INSDC(Genbank)} & \textbf{Refseqs} & \multicolumn{ 2}{c|}{\textbf{Input Cenomes}} \\ \hline
\textbf{Index } & \textbf{ GenomeName } & \textbf{ GID} & \textbf{ GID} & \textbf{GID} & \textbf{Accession no.} \\ \hline
1 &  PACS2  & 106896550 &  \_ & 106896550 & AAQW01000001.1 \\ \hline
2 &  PAO1  & 110227054 & 110645304 & 110645304 & NC\_002516.2 \\ \hline
3 &  UCBPP-PA14  & 115583796 & 116048575 & 116048575 & NC\_008463.1 \\ \hline
4 &  PA7  & 150958624 & 152983466 & 152983466 & NC\_009656.1 \\ \hline
5 &  19BR  & 343788106 & 485462089 & 485462089 & NZ\_AFXJ01000001.1 \\ \hline
6 &  213BR  & 343788107 & 485462091 & 485462091 & NZ\_AFXK01000001.1 \\ \hline
7 &  M18  & 347302377 & 386056071 & 386056071 & NC\_017549.1 \\ \hline
8 &  DK2  & 392316915 & 392981410 & 392981410 & NC\_018080.1 \\ \hline
9 &  B136-33  & 477548288 & 478476202 & 478476202 & NC\_020912.1 \\ \hline
10 &  RP73  & 514245605 & 514407635 & 514407635 & NC\_021577.1 \\ \hline
11 &  c7447m  & 543873856 &  \_ & 543873856 & CP006728 \\ \hline
12 &  PAO581  & 543879514 &  \_ & 543879514 & CP006705 \\ \hline
13 &  PAO1-VE2  & 553886202 &  \_ & 553886202 & CP006831 \\ \hline
14 &  PAO1-VE13  & 553895034 &  \_ & 553895034 & CP006832 \\ \hline
15 &  PA1  & 557703951 & 558672313 & 558672313 & NC\_022808.1 \\ \hline
16 &  PA1R  & 557709751 & 558665962 & 558665962 & NC\_022806.1 \\ \hline
17 &  MTB-1  & 563408818 & 564949884 & 564949884 & NC\_023019.1 \\ \hline
18 &  LES431  & 566561164 & 568151185 & 568151185 & NC\_023066 \\ \hline
19 &  SCV20265  & 567363169 & 568306739 & 568306739 & NC\_023149 \\ \hline
20 &  LESB58  &  \_ & 218888746 & 218888746 &  NC\_011770.1 \\ \hline
21 &  NCGM2.S1  &  \_ & 386062973 & 386062973 & NC\_017549.1 \\ \hline
22 &  PA38182  & 575870901 &  \_ & 575870901 & HG530068.1 \\ \hline
23 &  YL84  & 576902775 &  \_ & 576902775 & CP007147.1 \\ \hline
\end{tabular}}
\label{inputset}
\end{table}

Finally, in 2014, the analysis of the NGOs for all IS elements within 155 fully sequenced Archaea genomes was presented by Florek 
\emph{et al.}~\cite{art3florek2014insertion}. To do so,
they have launched a BLASTP in the ISFinder, with an e-value
less than or equal to $10^{-10}$, for 
all protein coding sequences downloaded from NCBI which are related to IS 
elements.


Two major concerns with the state of the art detailed above
can be emphasized. Firstly, most of them cannot detect new insertion sequences.
Secondly, all these tools are based on NCBI annotations of very relative
and variable qualities -- except ISsaga, which could work with other annotation tools (but it depends only on transposase ORFs that have been already defined in ISFinder). Our objective in the next section is to propose a pipeline 
that solves these two issues, being able to deal with unannotated genomes
and to detect unknown ISs.



\section{Prediction and Modules based on OASIS}
\label{sec:methods}

For illustration purpose, the proposed pipeline system for IS elements prediction will be
presented using 23 complete genomes of \emph{P. aeruginosa} available on the 
NCBI website, RefSeq and INCDS/Genebank databases, see Table~\ref{inputset}
(RefSeq genomes were prefered when available).
The prediction of IS elements in the proposed pipeline depends on both OASIS~\cite{robinson2012oasis} and ISFinder~\cite{siguier2006isfinder}.



\subsection{Prediction of IS elements from \emph{Pseudomonas aeruginosa}}


OASIS is used in this pipeline for predicting insertion sequences in prokaryotic
genomes. 
This latter detects ISs in each genome by finding conserved regions 
surrounding already-annotated transposase genes, which are identified by the 
word \emph{$¨$transposase$¨$} in the ``product'' field of the GenBank file. 
Obviously OASIS highly depends on the quality of annotations~\cite{robinson2012oasis}, while to determine whether a given gene is a transposase or not 
is a very difficult task (indeed transposases are among 
the most abundant and ubiquitous genes in nature~\cite{aziz2010transposases}, and they are widely separated in Prokaryote genomes). 
OASIS deals with files having genbank format. It takes 
them as input and then produces two output files for each provided genome.
The first one is a fasta file that contains all IS nucleotide sequences, 
with start and end positions. It also contains the amino acid sequence for each ORF.
The second file is a summary table providing attributes 
that describe the insertion sequence: set-id, family, group, 
IS positions, inverted repeat left (IRL) and right (IRR), 
and orientation. Remark that most of these information are 
in the ISFinder database too. Indeed  OASIS find them alone but
it extracts family names and group from ISFinder.

The main problem found in OASIS is solved in the proposed pipeline by 
using different types of annotations: NCBI will not be used alone, 
and gene functionality taken from annotation tools will either or not
be used depending on the situation. Finally, transposases 
within IS will be verified using ISFinder database.
OASIS can thus be used in two different ways in our pipeline, depending on the 
provided genbank file. These two modules have been named NOASIS, 
which  uses the original input genbank genome file provided by the NCBI (as it is, without any modification), and DOASIS, which deals with modified genbank files that have been updated to obtain more accurate results than NOASIS.  
These modules are described thereafter.

\subsection{Normal OASIS (NOASIS)}

For finding predicted IS in NOASIS module, we simply applied OASIS on the input 
set of genomes with their NCBI annotations,
that is, with the original downloaded genbank file. 
Using the reference genome named PAO1, the summary outputted by the pipeline is
given in Tables~\ref{pao1} 
and~\ref{pao2}. In these NOASIS tables, the summary produced by OASIS is enriched
with new features
described below:
\begin{itemize}
	\item{\textit{\textbf{Real IS}}}
	IS sequences that have best match (first hit) when using BLASTN with ISFinder database, an e-value equal to 0.0, and with a functionality of each ORF within the IS recognized as a transposase.
	\item{\textbf{\textit{Partial IS}}}
Sequences that match part of known IS from ISFinder (\emph{i.e.}, have e-value lower 
than $10^{-10}$) and have  also a transposase gene functionality for the ORFs.
    \item{\textit{\textbf{Putative New IS}}}
	Sequences with bad score after making a BLASTN with ISFinder, but 
	with a transposase.
	They may be real insertion sequences not already added in ISFinder database
	or false positives, requiring human curation. 
\end{itemize}

Applying this slightly improved version of OASIS in the 23 genomes
of \emph{Pseudomonas} leads to a major issue: surprisingly,
NOASIS found no real insertion sequences in some genomes 
like PACS2 or SCV20265. 
The problem is that OASIS find 
multiple copies of IS elements in each genome by identifying 
conserved regions surrounding transposase genes. However some 
of the considered genomes either have no information about transposase gene 
into their feature genbank tables or have simply no feature 
table in their genbank format files. This issue is at the 
basis of our improved module called DOASIS, which is explained below.

\begin{figure}[ht!]
\centering
\includegraphics[scale=0.25]{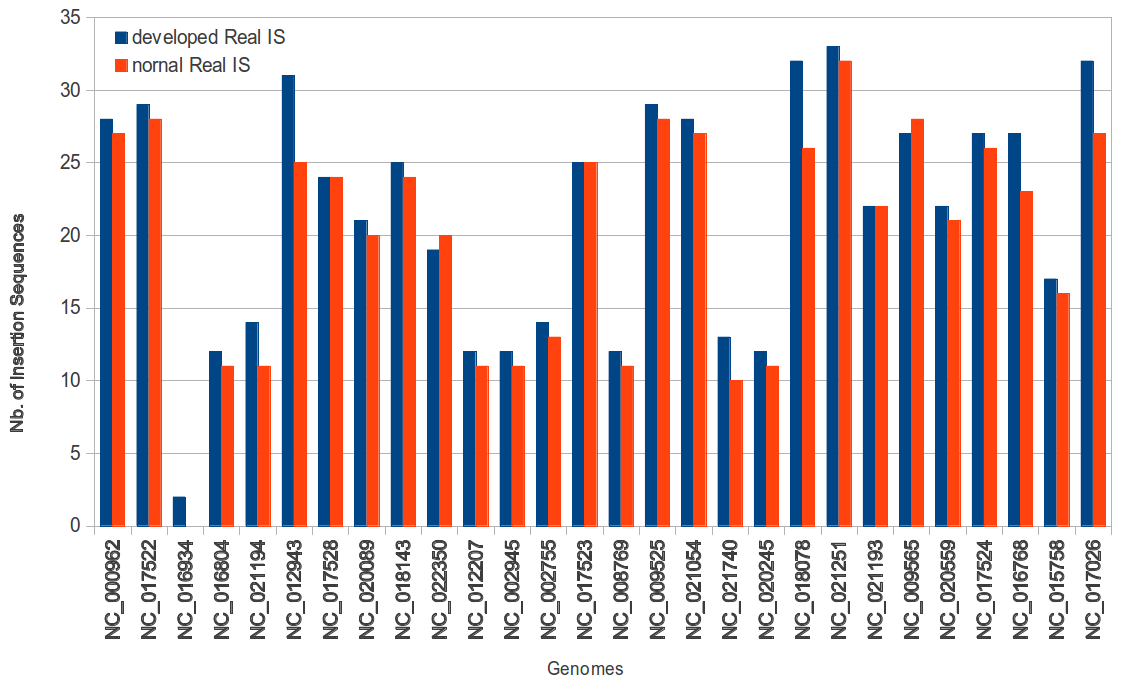}
\caption{IS elements detect in 28 \emph{Mycobacterium tuberculosis}}
\label{IS_tuber}
\end{figure}

For the sake of comparison, Figure \ref{IS_tuber} contains similar results for \emph{Mycobacterium tuberculosis} genus.
\color{black}

\subsection{Developed OASIS (DOASIS)}

The main idea for DOASIS module is that information about transposases within genbank files
are potentially incorrect (\emph{i.e.}, may all be false positives). So  we 
simply decide to remove all transposase words in the product fields 
from all inputted genomes. We thus update these information as follows.
\begin{description}
\item[Step 1: genbank update.] Inputted genbank files are modified following one of the three methods
 below.
\begin{enumerate}
\item \textbf{All-Tpase}: we consider that all the genes may potentially be a transposase.
So all product fields are set to ``transposase''.
\item \textbf{Zigzag Odd}: we suggest that genes in odd positions are putative transposases and we update the genbank file adequately. Oddly, this new path will produce new candidates which 
are not detected during All-Tpase.
\item \textbf{Zigzag Even}: similar to Zigzag Odd, but on even positions.
\end{enumerate}

We checked also a randomized method (\emph{i.e.}, by putting ``transposase'' in randomly picked genes). 
However we found poorer 
number of predictive real ISs or new real ISs compared with the three methods previously presented. 
For these reasons, we will not further investigate the randomized method.  

\color{black}

\begin{figure}[ht!]
\centering
\includegraphics[scale=0.4]{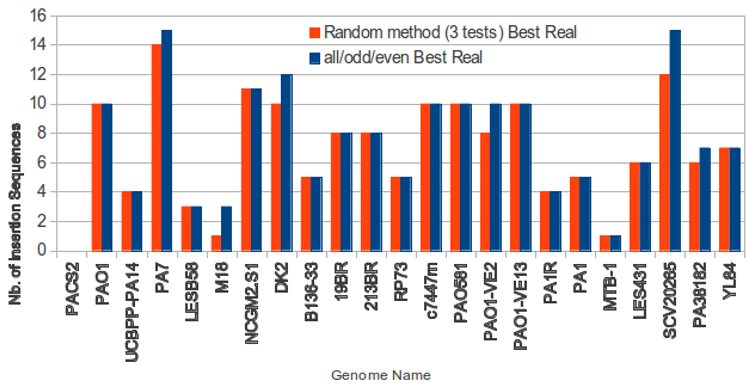}
\caption{Comparison of predicted ISs between randomization method and all/odd/even methods.}
\label{randomfig}
\end{figure}

\item[Step 2.] We apply OASIS three times (\emph{i.e.}, one time per method) 
on all genomes, and then
we take the output fasta file that 
contains both nucleotides and amino acids sequences for each IS element.
\item[Step 3.]  A BLASTN with ISFinder is applied on each IS sequence. 
If the e-value of the first hit is 0.0, then  the ORF within this 
IS belongs to known (Real) IS already existing in the ISFinder database. Else,
if the e-value is lower than $10^{-10}$, then we found a Partial IS.
\item[Step 4.] Collect all Real IS from previous three methods (ALL\_Tpase, Zigzag 
odd, and Zigzag even) and then remove overlaps among them. Finally, produce best 
Real IS with all information. Remark that the problem of finding consensus 
and overlaps can be treated as a lexical parsing problem.
\end{description}

\section {The Proposed Pipeline}
\label{sec:pip}

\begin{figure}[ht]
\centering
\includegraphics[scale=0.6]{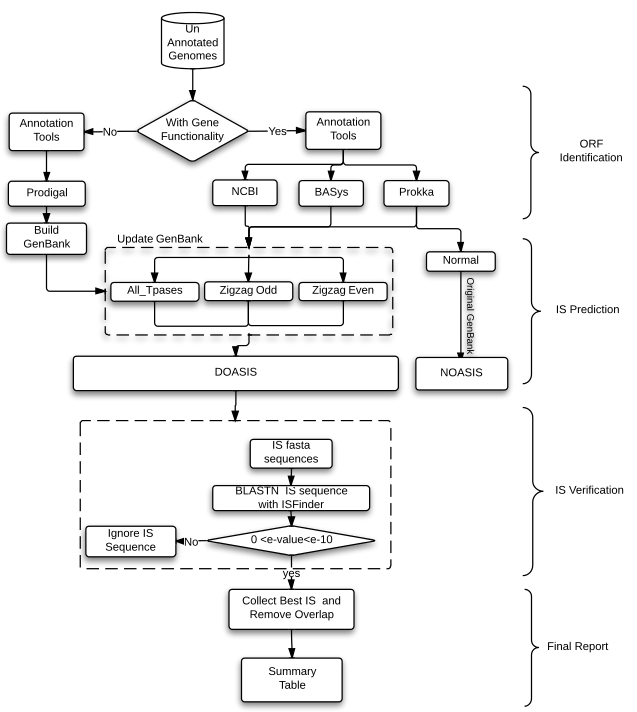}
\caption{The proposed pipeline}
\label{ISpip}
\end{figure}

It is now possible to describe the proposed pipeline that can use the two modules 
detailed in the previous section. This pipeline, depicted in Figure~\ref{ISpip}, 
will increase the number of
Real IS detected on the set of  \emph{P.aeruginosa} genomes under consideration 
(indeed, the detection is improved in all categories of insertion sequences,
but we only focus on Real IS in the remainder of this article, for the sake
of concision).
Its steps are detailed in what follows.


\begin{description}
\item[Step 1: ORF identification.]
Our pipeline is currently compatible with any type of annotation 
tools, having either functionality capability or not, but for comparison we only focus
in this article on the following tools: \emph{BASys}, \emph{Prokka}, and \emph{Prodigal}.
 BASys (Bacterial Annotation System) is a web server that performs 
automated, in-depth annotation of bacterial genomic (chromosomal and plasmid) 
sequences. 
It uses more than 30 programs to determine nearly 60 annotation subfields for each gene. 
Remark that genomes must be sent online manually, and that some 
curation stage may be required to remove some DNA ambiguity on 
returned genbank files.

Prokka (rapid prokaryotic genome annotation), for its part, is
a classical command line software for fully annotating draft bacterial genomes, 
producing standards-compliant output files for further analysis~\cite{seemann2014prokka}. 
Finally, 
Prodigal (Prokaryotic Dynamic Programming Genefinding Algorithm)
is an accurate bacterial and archaeal genes finding  software 
provided by the Oak Ridge National Laboratory~\cite{hyatt2010prodigal}.
\item[Step 2: IS Prediction.] The second stage of the pipeline consists in using 
either NOASIS or DOASIS for predicting IS elements. Notice that NOASIS 
cannot be used with Prodigal, as this module requires information
about gene functionality (both NOASIS and DOASIS can be use with 
Prokka and BASys annotations).
\item[Step 3: IS Validation.] This step is realized by launching BLASTN on each 
predicted IS sequence with ISFinder. The e-value of the first hit is then checked:
if it is 0.0, then the ORF within this sequence is a Real IS known by ISFinder. As 
described previously, it will be considered as Partial IS if 
its e-value is lower than $10^{-10}$. Both IS names of family and group are returned
too.
\end{description}

\section{Results and Discussion}
\label{sec:result}

\begin{figure}
\centering
\includegraphics[scale=0.5]{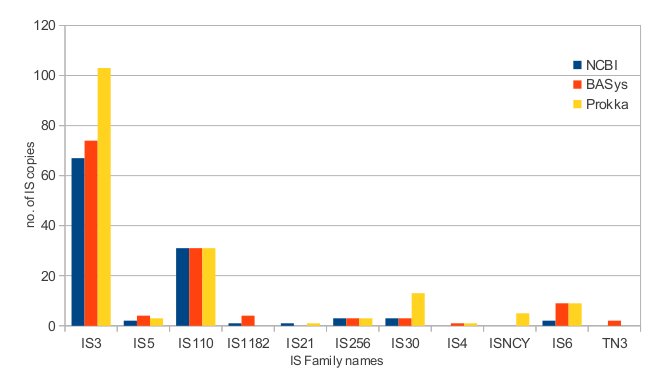}
\caption{Comparison between Prokka, BASys, and NCBI functionality annotations}
\label{fig:all_annot}
\end{figure}

We can firstly remark in Figure~\ref{fig:all_annot} that, 
using either Prokka or BASys for genes detection
and functionality prediction is better than taking directly the annotated genomes
from NCBI: a larger number of Real IS can be found. Additionally, this comparison
shows that Prokka outperforms BASys in 3 families of ISs (namely: IS3, IS30, and
ISNCY), while BASys seems better for detecting insertion sequences belonging
in the IS5, IS1182, and TN3 families. This variability may be explained by
the fact that functionality annotations of these tools depend probably on 
IS families that where known when these tools have been released.

\begin{table}[htbp]
\caption{Summary table produced by NOASIS (begining)}
\centering
\scalebox{0.7}{%
\begin{tabular}{|c|c|c|c|c|c|c|c|c|c|}
\hline
\textbf{Name} & \textbf{Genome} & \textbf{Start} & \textbf{End} & \textbf{Orientation} & \textbf{SetID} & \textbf{ISFinder\_name} & \textbf{Family} & \textbf{Group} & \textbf{Length} \\ \hline
PAO1 & NC\_002516.2 & 499832 & 501193 & - & 1 & ISPa11 & IS110 & IS1111 & 1361 \\ \hline
PAO1 & NC\_002516.2 & 2556875 & 2558236 & + & 1 & ISPa11 & IS110 & IS1111 & 1361 \\ \hline
PAO1 & NC\_002516.2 & 3043478 & 3044839 & - & 1 & ISPa11 & IS110 & IS1111 & 1361 \\ \hline
PAO1 & NC\_002516.2 & 3842002 & 3843363 & - & 1 & ISPa11 & IS110 & IS1111 & 1361 \\ \hline
PAO1 & NC\_002516.2 & 4473550 & 4474911 & + & 1 & ISPa11 & IS110 & IS1111 & 1361 \\ \hline
PAO1 & NC\_002516.2 & 5382524 & 5383885 & - & 1 & ISPa11 & IS110 & IS1111 & 1361 \\ \hline
PAO1 & NC\_002516.2 & 54041 & 54835 & + & 2 & ISStma5 & IS3 & IS3 & 794 \\ \hline
\end{tabular}}
\label{pao1}
\end{table}

\begin{table}[htbp]
\caption{Summary table produced by NOASIS (end)}
\centering
\scalebox{0.7}{%
\begin{tabular}{|c|c|c|c|c|}
\hline
\textbf{IRR=IRL} & \textbf{Locus\_tag(gbk)} & \textbf{Product(gbk)} & \textbf{E Value} & \textbf{IS\_type} \\ \hline
ATGGACTCCTCCC & [['PA0445']] & [['transposase']] & 0.0 & Real IS \\ \hline
ATGGACTCCTCCC & [['PA2319']] & [['transposase']] & 0.0 & Real IS \\ \hline
ATGGACTCCTCCC & [['PA2690']] & [['transposase']] & 0.0 & Real IS \\ \hline
ATGGACTCCTCCC & [['PA3434']] & [['transposase']] & 0.0 & Real IS \\ \hline
ATGGACTCCTCCC & [['PA3993']] & [['transposase']] & 0.0 & Real IS \\ \hline
ATGGACTCCTCCC & [['PA4797']] & [['transposase']] & 0.0 & Real IS \\ \hline
AAAGGGGACAGATTTATTTTCCCTGCTCTAAT & [['PA0041a']] & [['transposase']] & 0.23 & Putative New IS \\ \hline
\end{tabular}}
\label{pao2}
\end{table}

\begin{figure}[H]
\centering
\includegraphics[scale=0.35]{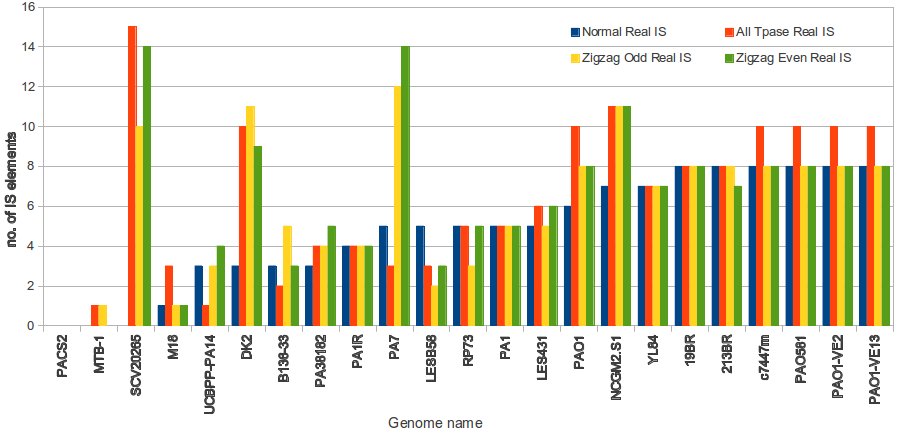}
\caption{NOASIS (NCBI annotation) versus DOASIS}
\label{fig:all_ncbi}
\end{figure}

The effects of DOASIS module compared to single OASIS
on annotated NCBI genomes are depicted in 
Figure~\ref{fig:all_ncbi}. The improvement in real IS
discovery is obvious, illustrating the low quality and inadequacy of 
NCBI annotations for studying insertion sequences in bacterial genomes,
and the improvements when using our pipeline. This chart
shows too that a zigzag path in the annotation can oddly
improve the detection of insertion sequences.

The prediction of real ISs is based on finding conserved regions (\emph{i.e.}, inverted repeats (IRs)) surrounded by transposase genes. 
Some ISs have been lost in All\_Tpase, for the following reason: when we suggested that all genes are transposases, OASIS found predicted ISs that consist of large sets of transposases surrounded by IR in their left and right boundaries. But when these predicted ISs have been verified using ISFinder database, we did not find any good match. Contrarily, in Zigzag methods, good matches have been found (real ISs), because many of these elements consist of one or two transposase genes flanked by IRs.
These results are listed with detail in Table~\ref{basys} using BASys annotation tools.

\begin{table}[H]
\caption{BASys annotation using NOASIS and DOASIS}
\begin{center}
\scalebox{0.6}{%
\begin{tabular}{|c|c|r|r|r|r|r|}
\hline
\textbf{BASys} &  & \multicolumn{1}{c|}{\textbf{Normal}} & \multicolumn{1}{c|}{\textbf{All Transpos}} & \multicolumn{1}{c|}{\textbf{Zigzag Odd}} & \multicolumn{1}{c|}{\textbf{Zigzag Even}} & \multicolumn{1}{c|}{\textbf{(All\_T/odd/even)}} \\ \hline
Name & Genome & \multicolumn{1}{l|}{Real IS} & \multicolumn{1}{l|}{Real IS} & \multicolumn{1}{l|}{Real IS} & \multicolumn{1}{l|}{Real IS} & \multicolumn{1}{c|}{\textbf{Best Real}} \\ \hline
PACS2 & 106896550 & 2 & 1 & 2 & 0 & 2 \\ \hline
PAO1 & 110645304 & 9 & 6 & 0 & 0 & 6 \\ \hline
UCBPP-PA14 & 116048575 & 3 & 8 & 8 & 1 & 8 \\ \hline
PA7 & 152983466 & 13 & 0 & 0 & 0 & 0 \\ \hline
LESB58 & 218888746 & 2 & 3 & 5 & 2 & 6 \\ \hline
M18 & 386056071 & 1 & 2 & 2 & 1 & 2 \\ \hline
NCGM2.S1 & 386062973 & 15 & 0 & 12 & 0 & 12 \\ \hline
DK2 & 392981410 & 8 & 9 & 10 & 8 & 11 \\ \hline
B136-33 & 478476202 & 3 & 5 & 3 & 3 & 5 \\ \hline
19BR & 485462089 & 5 & 0 & 0 & 10 & 10 \\ \hline
213BR & 485462091 & 5 & 4 & 4 & 4 & 4 \\ \hline
RP73 & 514407635 & 4 & 5 & 5 & 2 & 5 \\ \hline
c7447m & 543873856 & 9 & 0 & 9 & 9 & 10 \\ \hline
PAO581 & 543879514 & 9 & 6 & 8 & 0 & 8 \\ \hline
PAO1-VE2 & 553886202 & 8 & 6 & 9 & 6 & 9 \\ \hline
PAO1-VE13 & 553895034 & 8 & 6 & 8 & 8 & 8 \\ \hline
PA1R & 558665962 & 4 & 4 & 4 & 5 & 5 \\ \hline
PA1 & 558672313 & 5 & 5 & 5 & 6 & 6 \\ \hline
MTB-1 & 564949884 & 0 & 1 & 0 & 1 & 1 \\ \hline
LES431 & 568151185 & 5 & 14 & 13 & 8 & 14 \\ \hline
SCV20265 & 568306739 & 5 & 14 & 13 & 8 & 14 \\ \hline
PA38182 & 575870901 & 1 & 3 & 1 & 2 & 4 \\ \hline
YL84 & 576902775 & 7 & 7 & 7 & 7 & 7 \\ \hline
 &  & \multicolumn{1}{c|}{131} & \multicolumn{1}{c|}{109} & \multicolumn{1}{c|}{128} & \multicolumn{1}{c|}{91} & \multicolumn{1}{c|}{157} \\ \hline
\end{tabular}}
\end{center}
\label{basys}
\end{table}

We can thus wonder if the source of a wrong prediction of
real IS is due to a wrong coding sequence prediction, or to 
functionality errors. Switching between NOASIS and DOASIS
allows us to answer this question. We can conclude 
from Table~\ref{correlation} that (1) annotation errors are more frequent on NCBI, while Prokka annotates well the sequences related to ISs (see NOASIS columns), and that (2) both NCBI and Prokka have a better coding sequence prediction than BASys, at least when considering sequences involved in IS elements (see DOASIS columns and the correlation line). More precisely, the correlation is based on the number of predicted real IS elements between NOASIS and DOASIS.

\begin{table}[H]
\caption{Correlation table for different annotation tools} 
\begin{center}
\scalebox{0.7}{%
\begin{tabular}{|c|c|c|c|c|c|c|}
\hline
 \textbf{} & \multicolumn{ 2}{c|}{\textbf{NCBI}} & \multicolumn{ 2}{c|}{\textbf{BASys}} & \multicolumn{ 2}{c|}{\textbf{Prokka}} \\ \hline
 & \textbf{NOASIS} & \textbf{DOASIS} & \textbf{NOASIS} & \textbf{DOASIS} & \textbf{NOASIS} & \textbf{DOASIS} \\ \hline
 Number of Real IS & 110 & 169 & 131 & 157 & 169 & 176 \\ \hline
 \textbf{Correlations} & \multicolumn{ 2}{c|}{0.3985580752} & \multicolumn{ 2}{c|}{0.357346472} & \multicolumn{ 2}{c|}{0.926355615} \\ \hline
\end{tabular}}
\end{center}
\label{correlation}
\end{table}

\begin{figure}[h]
\centering
\includegraphics[scale=0.5]{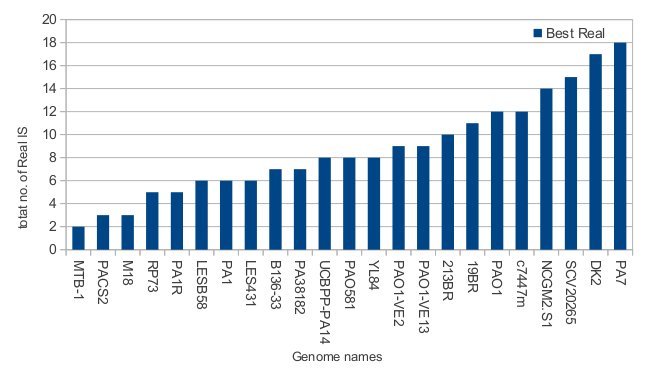}
\caption{Real ISs found by our pipeline using Prodigal}
\label{prod}
\end{figure}

Prodigal has been studied separately, as it does not provide genes functionality.
The number of Real ISs per genome returned by our pipeline
using prodigal is given in Figure~\ref{prod}. As shown in Table~\ref{high throughput},
the quality of  coding sequences predicted with prodigal compared
with other annotation tools allows us to discover the best number of real ISs. 
In particular, we have improved a lot of results produced
by OASIS and ISFinder on NCBI annotations, which is usually used in the
literature that focuses on bacterial insertion sequences.
Furthermore, this table illustrates a certain sensitivity of 
coding sequence prediction tools with functionality annotation capabilities
to detect ISs in some specific genomes like PA7.  
Indeed we discovered, 
during other studies we realized on this set of \emph{Pseudomonas} strains,
that PA7 has a lot of specific genes, that is, which are not in the core genome of all \emph{Pseudomonases}, which may explain such a sensitivity. 

\begin{table}[htbp]
\caption{Final comparison using our pipeline}
\centering
\scalebox{0.7}{%
\begin{tabular}{|c|c|c|c|c|c|c|c|c|c|}
\hline
\textbf{} & \textbf{NCBI} & \textbf{BASys} & \textbf{Prokka} & \textbf{Prodigal} & \textbf{} & \textbf{NCBI} & \textbf{BASys} & \textbf{Prokka} & \textbf{Prodigal} \\ \hline
\textbf{Name} & \textbf{DOASIS} & \textbf{DOASIS} & \textbf{DOASIS} & \textbf{DOASIS} & \textbf{Name} & \textbf{DOASIS} & \textbf{DOASIS} & \textbf{DOASIS} & \textbf{DOASIS} \\ \hline
PACS2 & 0 & 2 & 2 & 3 & c7447m & 10 & 10 & 10 & 12 \\ \hline
PAO1 & 10 & 6 & 10 & 12 & PAO581 & 10 & 8 & 10 & 8 \\ \hline
UCBPP-PA14 & 4 & 8 & 4 & 8 & PAO1-VE2 & 10 & 9 & 10 & 9 \\ \hline
PA7 & 15 & 0 & 14 & 18 & PAO1-VE13 & 10 & 8 & 10 & 9 \\ \hline
LESB58 & 3 & 6 & 3 & 6 & PA1R & 4 & 5 & 4 & 5 \\ \hline
M18 & 3 & 2 & 3 & 3 & PA1 & 5 & 6 & 5 & 6 \\ \hline
NCGM2.S1 & 11 & 12 & 19 & 14 & MTB-1 & 1 & 1 & 1 & 2 \\ \hline
DK2 & 12 & 11 & 13 & 17 & LES431 & 6 & 14 & 6 & 6 \\ \hline
B136-33 & 5 & 5 & 5 & 7 & SCV20265 & 15 & 14 & 15 & 15 \\ \hline
19BR & 8 & 10 & 5 & 11 & PA38182 & 7 & 4 & 7 & 7 \\ \hline
213BR & 8 & 4 & 8 & 10 & YL84 & 7 & 7 & 7 & 8 \\ \hline
RP73 & 5 & 5 & 5 & 5 & \textbf{Total IS} & \textbf{84} & \textbf{71} & \textbf{91} & \textbf{114} \\ \hline
\end{tabular}}
\label{high throughput}
\end{table}

\section {Conclusion}
\label{sec:conclusion}

Insertion sequences of bacterial genomes are usually studied using
OASIS and ISFinder on NCBI annotations. We have shown in this 
article that a pipeline can be designed to improve the accuracy
of IS detection and classification by improving the coding sequence
prediction stage, and by considering a priori each sequence as a transposase. The source code for this pipeline can be download from the link~\footnote{\url{http://members.femto-st.fr/christophe-guyeux/en/insertion-sequences}}. 
A comparison has been conducted on a set of \emph{Pseudomonas aeruginosa},
showing an obvious improvement in the detection of insertion
sequences for some particular configurations of our pipeline.

In future work, we intend to enlarge the number of coding sequence
and functionality prediction tools and to merge all the Real IS results
in order to improve again the accuracy of our pipeline. We will then focus
on the impact of IS elements in \emph{P.aeruginosa} evolution,
comparing the phylogenetic tree of strains of this species with 
a phylogeny of their insertion sequences. Insertion events will then
be investigated, and related to genomes rearrangements found in this
collection of strains. We will finally enlarge our pipeline to 
eukariotic genomes and to other kind of transposable elements.

\bibliographystyle{plain}
\bibliography{biblio}

\end{document}